\documentclass[twocolumn]{aastex62}

\newcommand{\msun}{M_{\odot}}
\newcommand{\sigvbc}{\sigma_{v_{\rm bc}}}
\newcommand{\vbc}{v_{\rm bc}}
\newcommand{\ls}{\lambda_{\rm s}}

\newcommand{\lj}{\lambda_{\rm J}}
\newcommand{\cs}{c_{\rm s}}
\usepackage{color}

\received{XX}
\revised{YY}
\accepted{ZZ}

\submitjournal{ApJL}

\shorttitle{SIGO luminosity}
\shortauthors{Chiou et al.}

\begin{document}

\title{The Supersonic Project: Shining Light on SIGOs -  a New Formation Channel for Globular Clusters  }

\correspondingauthor{Yeou S. Chiou}
\email{yschiou@physics.ucla.edu}

\author{Yeou S. Chiou}
\affil{Department of Physics and Astronomy, University of California, Los Angeles, CA 90095\\}
\affil{Mani L. Bhaumik Institute for Theoretical Physics, Department of Physics and Astronomy, UCLA, Los Angeles, CA 90095, USA\\}

\author{Smadar Naoz}
\affil{Department of Physics and Astronomy, University of California, Los Angeles, CA 90095\\}
\affil{Mani L. Bhaumik Institute for Theoretical Physics, Department of Physics and Astronomy, UCLA, Los Angeles, CA 90095, USA\\}

\author{Blakesley Burkhart}
\affiliation{Department of Physics and Astronomy, Rutgers, The State University of New Jersey, 136 Frelinghuysen Rd, Piscataway, NJ 08854, USA \\}
\affiliation{Center for Computational Astrophysics, Flatiron Institute, 162 Fifth Avenue, New York, NY 10010, USA \\}

\author{Federico Marinacci}
\affiliation{Department of Physics \& Astronomy, University of Bologna, via Gobetti 93/2, 40129 Bologna, Italy\\}
\affiliation{Harvard-Smithsonian Center for Astrophysics, 60 Garden Street, Cambridge, MA 02138, USA\\}

\author{Mark Vogelsberger}
\affil{Department of Physics and Kavli Institute for Astrophysics and Space Research, Massachusetts Institute of Technology, Cambridge, MA 02139, USA\\}

\begin{abstract}

Supersonically induced gas objects (SIGOs) with little to no dark matter component are predicted to exist in patches of the Universe with non-negligible relative velocity between baryons and the dark matter at the time of recombination. Using {\sc arepo} hydrodynamic simulations we find that the gas densities inside these objects are high enough to allow stars to form. An estimate of the luminosity of the first star clusters formed within these SIGOs suggests that they may be observed at high redshift using future HST and JWST observations. Furthermore, our simulations indicate that SIGOs lie in a distinct place in the luminosity-radius parameter space, which can be used observationally to distinguish SIGOs from dark-matter hosting gas systems. Finally, as a proof-of-concept, we model star formation before reionization and evolve these systems to current times. We find that SIGOs occupy a similar part of the magnitude-radius parameter space as globular clusters. These results suggest that SIGOs may be linked with present-day metal-poor local globular clusters. Since the relative velocity between the baryons and dark matter is coherent over a few Mpc scales, we predict that if this is the dominant mechanism for the formation of globular clusters, their abundance should vary significantly over these scales.   

\end{abstract}

\keywords{cosmology: theory -- methods: numerical -- galaxies: high redshift}

\section{Introduction}\label{sec:Intro}

The puzzling origins of globular clusters (GCs) have been greatly debated over the years \citep{gunn80,peebles84,ashman+92, harris+94,grillmair+95,moore96,Bromm+02b,kravtsov+05,mashchenko+05,saitoh+06,muratov+10,bekki+12,kruijssen15, renaud+17,mandelker+18}. These objects serve as the testing grounds for early structure formation since they are very old  \citep[$\sim13$~Gyr, e.g.,][]{Trenti+15}. For example, they have even been used to estimate the age of the Universe \citep{Krauss+03}. Observations suggest that GCs contain practically no gravitationally bound dark matter \citep[e.g.,][although see \citet{taylor+15} for evidence to the contrary]{Heggie+96,Bradford+11,Conroy+11,Ibata+13}. Although direct observations of high redshift GCs is difficult, statistical studies with strong gravitational lensing have enabled the investigation of high redshift star-forming GC candidates   \citep[e.g.,][]{elmegreen+12, vanzella+17}. There has even been some direct evidence of possible GC progenitors \citep[e.g.,][]{vanzella+16,bouwens18,vanzella+19}. Furthermore, GCs and their progenitors may also play a large role in reionizing the Universe \citep[e.g.,][]{schaerer+11,BK18}. The upcoming \textit{James Web Space Telescope} (JWST) offers an exciting chance to observe GCs and their progenitors at early times. These observations will give insight to the formation of the very early building blocks in the Universe. 

In the standard model of structure formation, due to the baryon-radiation coupling, baryon over-densities at the time of recombination ($z\sim 1020$) were about 5 orders of magnitude smaller than dark matter (DM) over-densities. \citet{TH} showed that not only were the amplitudes of the DM and baryonic density fluctuations different at early times \citep[e.g.,][]{NB}, but so were their velocities. After recombination, the baryons decoupled from the photons and their subsequent evolution was dominated by the gravitational potential of the DM. In the period following recombination, the baryons underwent rapid cooling. At this point, their relative velocity with respect to the DM, which at recombination was of the order of $\sim 30$~km~sec$^{-1}$, became supersonic. \citet{TH} also showed that this relative velocity between the baryons and the DM remained coherent on scales of a few Mpc and in these regions it can be modeled as a stream velocity. 

The stream velocity effect has previously been overlooked because the velocity terms are formally second order in perturbation theory and are therefore neglected in the linear approximation. However, this second-order effect is unusually large, resulting in the non-negligible suppression of power at mass scales that correspond to the first bound objects in the Universe \citep{tseliakhovich11}. The non-linear effects of the stream velocity on the first structures were subsequently investigated using numerical simulations \citep[e.g.,][]{Stacy+10,miao11,greif11,fialkov2012,naoz11,naoz12,oleary12,richardson13,tanaka14}. The stream velocity also has significant implication on the redshifted cosmological 21-cm signal \citep[e.g.][]{dalal10,Visbal+12,McQuinn+12}, the formation of primordial black holes \citep[e.g.,][]{tanaka13,tanaka14,latif14,hirano17,schauer17}, and even for primordial magnetic fields \citep{naozyoshida13}. See  \citet{Fialkov14} for a detailed review. 

Recently, \citet{naoznarayan14} proposed that metal-poor GCs may be linked to objects that can be formed without DM in the early Universe in the presence of the stream velocity. These supersonically induced gas objects (SIGOs) were later found in numerical simulations by \citet{popa,chiou18}. However, their connection to GCs is still uncertain. Specifically, the ability of SIGOs to form stars was not addressed in those simulations. If these objects indeed form stars, these first star clusters will host little to no DM component.

The formation of the first stars from pristine gas was addressed in length in the literature focusing on the detailed chemistry and equation of state \citep[e.g.,][]{Abel+02,Bromm+02b,Reed+05,Yoshida+06,Stacy+10,Glover13,Xu+16,Sarmento+18,schauer+19}. In this {\it letter} we take a global, statistical approach through an investigation of the conditions for star formation in SIGOs via a simple density threshold argument. In particular, stars will form if the global gas density within a given SIGO is above the predicted critical value for star formation in pristine and low-metallicity gas \citep[e.g.,][]{Christlieb+02,Krumholz+05,burkhart18b}. We then use semi-analytical calculations to determine the luminosities of objects in our simulations. We note that, although we study these objects at $z=20$, they still exists at lower redshifts \citep[e.g.,][]{naoznarayan14,popa,chiou18}. Thus, their expected luminosities could possibly be detectable with JWST. 

The {\it letter} is organized as follows: we begin by describing our simulations in Section \ref{sec:OC} then we discuss the star formation model (Section  \ref{sec:StarForm}) and  the subsequent luminosity (Section \ref{sec:Results}). Finally we offer our discussion and qualitative predictions in Section  \ref{sec:Conclusions}.

\section{Simulation details and object classification}\label{sec:OC}

 We run two cosmological simulations with the moving-mesh code {\sc arepo} \citep{springel10} in a $2$~Mpc box with $512^3$ DM particles of mass $M_{\rm DM} = 1.9 \times 10^3 \msun$ and $512^3$ Voronoi mesh cells with $M_{\rm gas}=360 \msun$. One run has a stream velocity value of $\vbc=~2\sigvbc$, where $\sigvbc$ is the rms value of the stream velocity (i.e., the relative velocity between the gas and the dark matter component), while the other, which we use for comparison, has no such stream velocity. Both runs include radiative cooling \citep[see][]{chiou+19prep}. The cooling module in {\sc arepo} is based on a self-consistent primordial chemistry and cooling network, which includes the evolution of species H, H$^+$, He, He$^+$, He$^{++}$ and e$^-$ in equilibrium with a photoionizing background that is spatially constant but redshift dependent. The gas cooling and heating rates are calculated as a function of redshift, gas density, temperature and (for the metal line part) metallicity\footnote{Note that while all cooling rates include self-shielding corrections, these corrections do not apply above redshift of $6$, and thus do not contribute for the cooling of the $z=20$ objects.} See \citet[][and references therein]{vogelsberger+13} for further details on the numerical implementation of these processes. The simulations do not include explicit star formation or feedback.
 
We note that our simulations do not include H$_2$ cooling. Molecular cooling was shown to be  important for early star formation \citep[e.g.][]{Hartwig+15,Glover+07,schauer17,schauer18}. Nonetheless, the densities in our simulations (as we show below) reach the necessary high densities and low temperatures to trigger star formation. Thus, inclusion of molecular cooling in followup simulations will yield even higher densities, further facilitating star formation.    

The initial conditions adopted different transfer functions for the DM and baryon components as described in \citet{NB,Naoz+09,naoz11,naozyoshida13,naoz12}. The runs were performed from a redshift of $z=200$ to $z=20$. The stream velocity is implemented as a uniform boost for the gas in the $x$-direction. The choice of $v_{\rm bc}=2\sigma_{v_{\rm bc}}$ allows us to gain a larger effect, however the same physical picture is applicable for $v_{\rm bc}=1\sigma_{v_{\rm bc}}$, \citep[as noted by][]{naoznarayan14}. Furthermore, we gain more statistical power by adopting $\sigma_8=1.7$ \citep[e.g.,][]{popa,chiou18,chiou+19prep}. 

We follow the structure definitions suggested in \citet{chiou18}. In particular, DM-Primary/Gas-Secondary (DM/G) objects are spherical overdensity DM halos that also contain gas. The Gas-Primary objects are 
gas objects obtained through running a Friends-of-Friends (FOF) algorithm on only the gas component and subsequently fitted to a tight ellipsoid. Both DM/G and Gas-Primary objects are identified by using the FOF algorithm with a linking length of $0.2$ times the mean particle separation. Finally, the SIGOs are Gas-Primary objects that are outside the virial radius of the closest DM halo and also have gas fractions greater than $40\%$. These objects have little to no DM component. The advantage of our small simulation box allows us to resolve SIGOs, however, it prevents us from following the detailed evolution of SIGOs to smaller redshift. Thus, in order to investigate the evolution of SIGOs for $z<20$ we employ semi-analytical modeling. Throughout the paper we assume a $\Lambda$CDM cosmology with $\Omega_{\Lambda} = 0.73$, $\Omega_m= 0.27$, $\Omega_B= 0.044$, $\sigma_8= 1.7$, and $h = 0.71$. All the quantities that we analyze in this paper are expressed in physical units.

\begin{figure*}
\includegraphics[width=.99\textwidth]{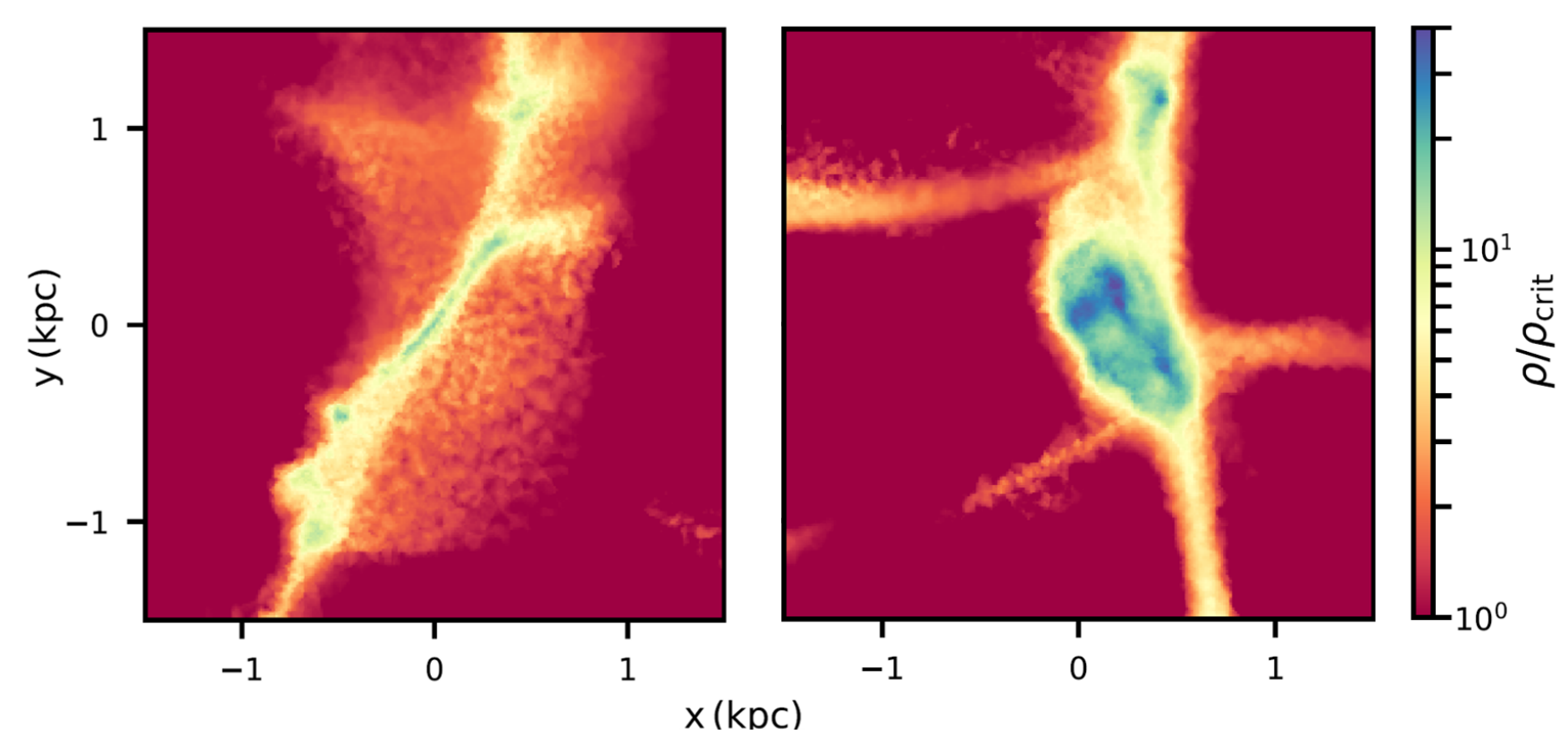}
\caption{Density projections of two representative SIGOs at $z=20$. The density has been normalized to $\rho_{\rm crit}$ for each SIGO. The highest density ratio here is $\sim 40$, corresponding to dark blue color. Note the filamentary nature with multiple high, smaller scale density peaks. The SIGO is embedded at the center of the region with a scale of ($R_{\rm min}, R_{\rm max}$) of (.01, .03) kpc and (.08, .14) kpc for the left and right panels, respectively.}
\label{fig:densproj}
\end{figure*}

\section{Star formation model}
\label{sec:StarForm}

We estimate the plausibility that a dense gas clump (either a SIGO or within a DM/G) may form stars. Primordial star formation may be the 
most suitable epoch during the evolution of the Universe for the application of the Jeans criterion since the level of turbulence and strength of the magnetic field are expected to be significantly lower \citep{bromm+99,bromm+02a}. The Jeans criterion \citep{jeans1902} and the related Bonnor-Ebert mass describes the balance between gravity and thermal pressure and is given by:
\begin{equation}\label{eq:MBE}
M_{\rm BE}  =  1.18 \frac{\cs^3}{\sqrt{G^3 \rho}} =  \frac{1.18}{\pi^{3/2}} \rho \lj^3 \ ,
\end{equation}
where $\cs$ is the isothermal sound speed in the region, $G$ is the gravitational constant, $\rho$ is the density of the gas, and $\lj$ the Jeans length.  
The mass in Equation (\ref{eq:MBE})  is the largest mass that an isothermal gas sphere embedded in a pressurized medium can have while still remaining in hydrostatic equilibrium \citep{Ebert55a,Bonnor56a}.  This depends on the Jeans length, defined as follows:
\begin{equation}
\lj = \sqrt{\frac{\pi \cs^2}{G\rho}} \ .
\end{equation}
The Jeans length is therefore the critical radius of a cloud where thermal energy is counteracted by gravity.

Since we deal with a supersonic medium the other length scale of interest for defining a critical density for star formation is the sonic scale\footnote{This expression assumes a line-width size relation with exponent of p=0.5, expected for supersonic turbulence}:
\begin{equation}
\label{linewidth2}
\ls= \left(\frac{L_{\rm drive}}{\mathcal{M}^2}\right) \ .
\end{equation}
 $\ls$ is defined as the length scale such that $\sigma_l=\cs$,
 $\mathcal{M}$ is the mach number on the driving scale, $L_{\rm drive}$, of the turbulence, and $\sigma_l$ is the one-dimensional velocity
dispersion computed over a sphere of diameter $l$ within a turbulent medium. The sonic scale physically represents the scale at which turbulence in the gas transitions from supersonic to subsonic.

As discussed in \citet{Krumholz+05}, if $\lj \le \ls$,
gravity is approximately balanced by thermal plus turbulent pressure, and the object is at best marginally stable against collapse. Here we assume that the magnetic field is dynamically unimportant relative to turbulence and gravity. If $\lj\gg\ls$, turbulent/thermal kinetic energy greatly exceeds gravitational potential energy and the object is stable against collapse. 
Since $\lj$ is a function of the local density, the condition $\lj \le \ls$ for collapse translates into a minimum local density
required for collapse (in the absence of magnetic fields).
Equating the two length scales yields a critical density  
\begin{equation}\label{eq:rhocrit}
\rho_{\rm crit} = \frac{\pi c_s^2 \mathcal{M}^4}{GL_{\rm drive}^2} \ ,
\end{equation}
 which can be rewritten in terms of the virial parameter, and assuming that the driving scale of the turbulence is the characteristic diameter of the cloud, i.e., $L_{\rm drive}=L_{\rm cloud}$, as,
 \begin{equation}\label{eq:rhocrit2}
\rho_{\rm crit} = \frac{\pi^2}{15}\rho_0 \alpha_{\rm vir,l}\mathcal{M}^2 \ ,
\end{equation}
 where  $\alpha_{\rm vir,l}= {5\sigma_l^2L_{\rm cloud}}/({2G M_{\rm cloud}})$, the ratio of turbulent to gravitational energy. We note that these equations are only meaningful in the presence of a supersonic flow, as in our case.
 
In this simple stability picture, once the critical density is reached, the gas becomes unstable to gravitational collapse. \citet{burkhart18b} related the critical density to a transition density between a piecewise lognormal and power law of the density distribution. A power-law density PDF is the 1-point statistic's signature of gravitational collapse, regardless of the gas metallicity.  A turbulent medium will have an initially lognormal density distribution, but once gravitational collapse sets in, the distribution can be described by a power law \citep{girichidis+14,burkhart+17,guszejnov+18}. Thus, the transition density as a critical density for collapse is a natural consequence of the density distribution function. Below, we adopt this critical density threshold as a star formation indicator.

\begin{figure}
    \centering
    \includegraphics[width=.5\textwidth]{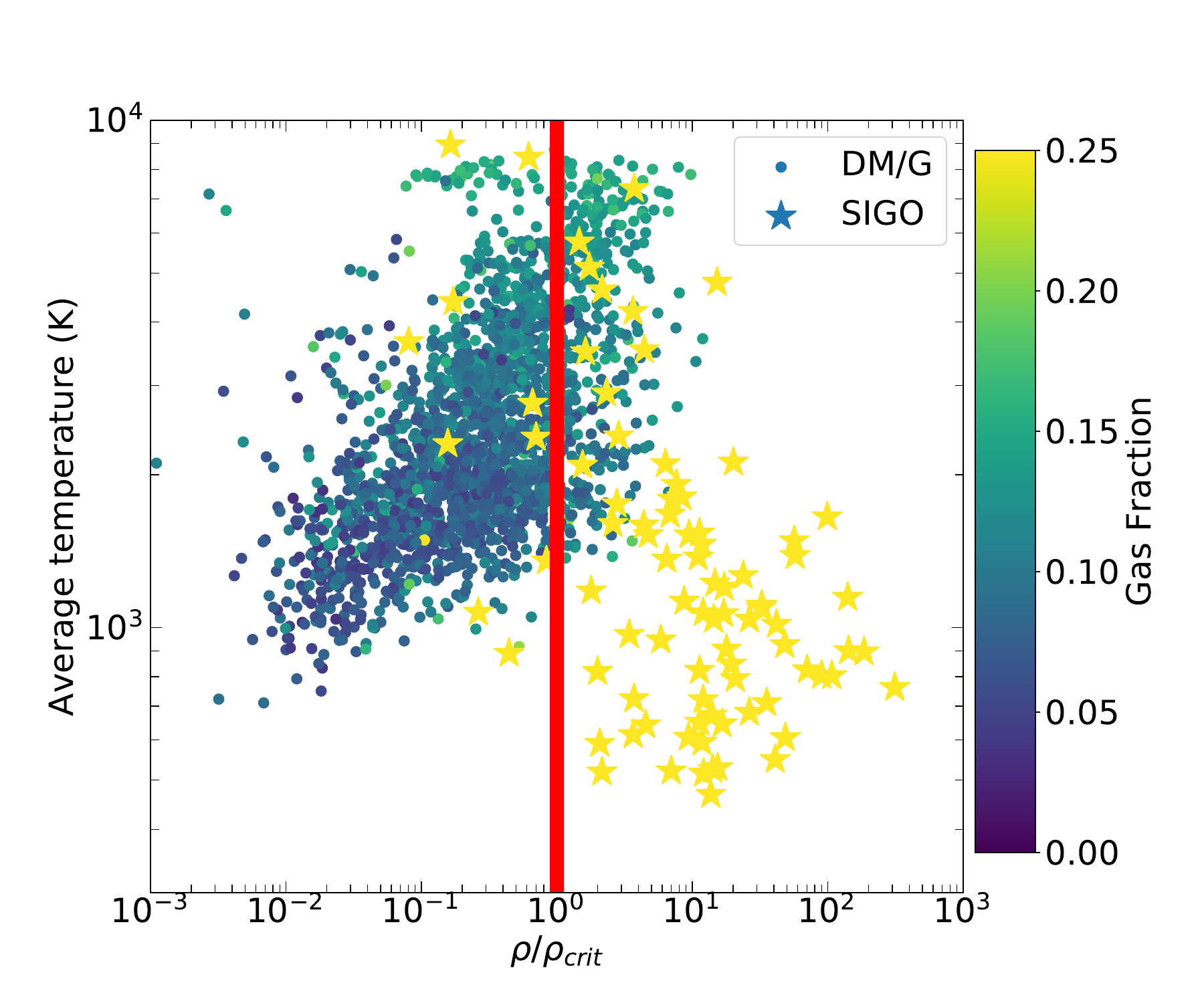}
    \caption{The average temperature in the stream velocity run  as a function of $\rho/\rho_{\rm crit}$, at $z=20$. The red vertical line indicates the $\rho=\rho_{\rm crit}$ line. To the right of the line we expect star formation to take place and to the left, star formation is suppressed. The color code depicts the gas fraction in the object. Recall that by definition SIGOs have gas fraction above $40\%$. Note the majority SIGOs are in the star forming regime.     } 
    \label{fig:tempvsrho}
\end{figure}

We apply the above star formation criterion to the objects found in the simulation. In Figure \ref{fig:densproj}, we show the density of representative star forming SIGOs normalized to their critical density for star formation.   
{To compute the critical density we need an estimate of the turbulence sonic scale, which} is given by $L_{\rm cloud}/\mathcal{M}^2$ \citep[e.g.,][]{burkhart18a}. Since SIGOs are ellipsoidal \citep[e.g.,][]{chiou18}, we assume $L_{\rm cloud} \sim 2R_{\rm max}$, because this is the maximum scale at which turbulence can be generated. 
 We calculate the critical densities for each object type (i.e., DM/Gs and SIGOs) in our simulations following Equation (\ref{eq:rhocrit}). Considering first the DM/G objects we find that at $z=20$, $19\%$ ($85\%$) of them have densities and temperatures that yield favorable conditions for star formation for the $2\sigvbc$ ($\vbc=0$) run. This difference between the stream velocity and no stream velocity case is expected since the stream velocity effect reduces the gas fraction of DM halos \citep[e.g.,][]{tseliakhovich11, naoz12}. In Figure \ref{fig:tempvsrho} we show the temperatures and densities for the DM/G objects in the presence of stream velocity.  As expected DM halos that host larger gas fractions are more likely to form stars, according to the $\rho_{\rm crit}$ criterion.  Note that gas in the DM/G objects is expected to fragment into cooler clumps that will serve as star formation sites \citep[e.g.,][]{bromm+99,greif11}. 

Significantly, $88\%$ of SIGOs may form stars for the $2\sigvbc$ run (there are ipso facto no SIGOs in the $\vbc=0$ run).  As depicted in Figure \ref{fig:tempvsrho}, SIGOs have densities that are much higher than $\rho_{\rm crit}$ and are overall cool. {\it  In other words, the majority of SIGOs  have supercritical densities and are thus ripe sites for star formation.}

Note that since SIGOs are only marginally bound \citep{chiou18}, supernova feedback may disrupt the rest of the gas in them, thus suppressing further star formation. Indeed GCs tend to have multiple generations of stars, possibly from multiple star burst epochs. {Subsequent star bursts may form during pericenter passage as a SIGO orbits the closest DM halo, if gas survived the supernova feedback or was able to accrete from the medium\footnote{Here we adopt a semi-analytical approach for star formation, since detailed zoom-in simulations exploring the supernova feedback are beyond the scope of this letter.}.} Here we focus on the first star formation episode and hence adopt a star burst formation model.

 \begin{figure} 
\includegraphics[width=.45\textwidth]{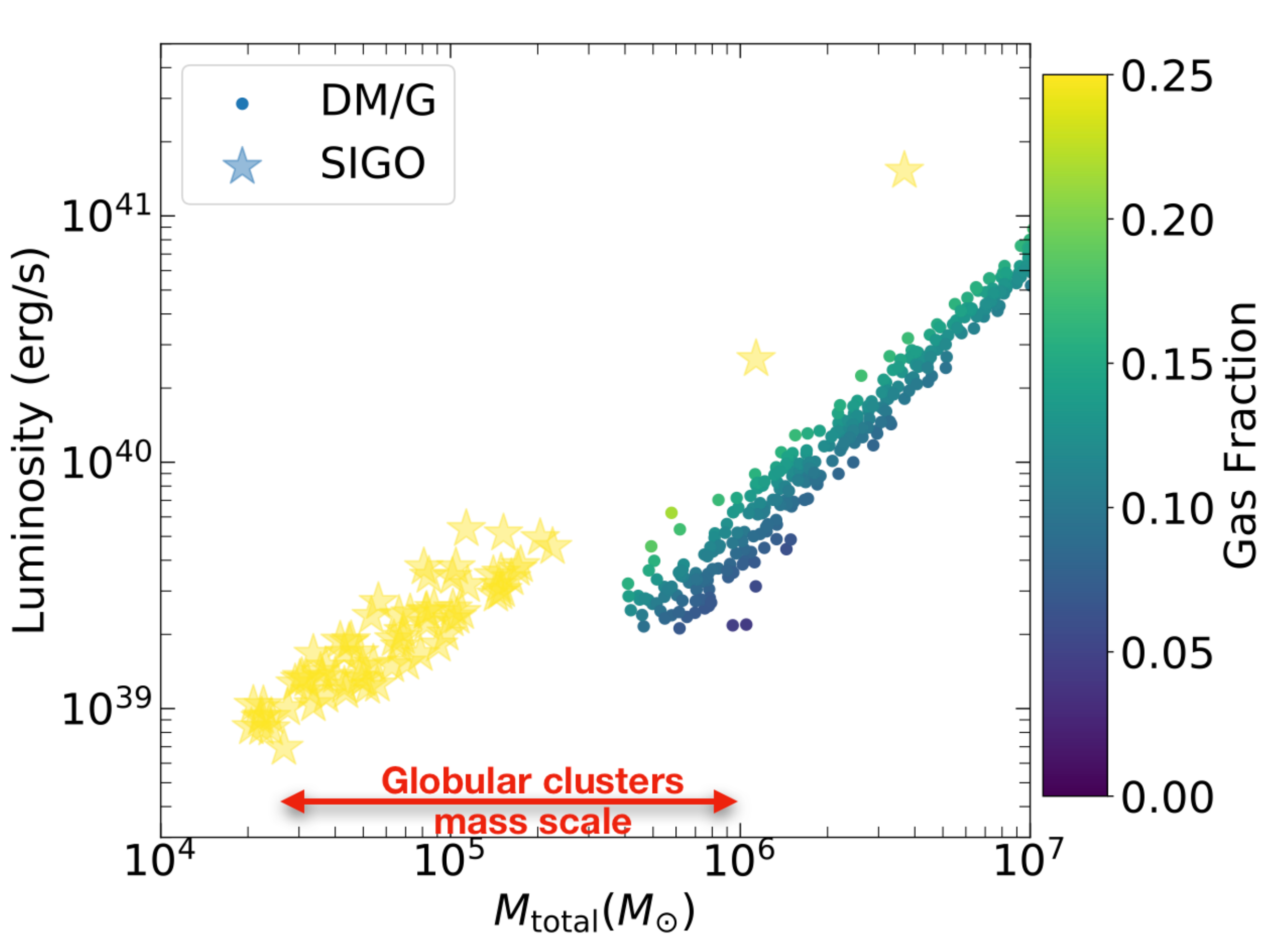}
\caption{Luminosity as a function of mass for the $v_{\rm bc} = 2\sigma_{v_{\rm bc}}$ run at $z=20$. The color code depicts the gas fraction in the object. The horizontal line represents a characteristic mass scale of present-day, local, globular clusters \citep[e.g.,][]{Kimmig+15}.}
\label{fig:lumvsmass}
\end{figure}

\begin{figure*} 
\begin{center}
\includegraphics[width=.90\textwidth]{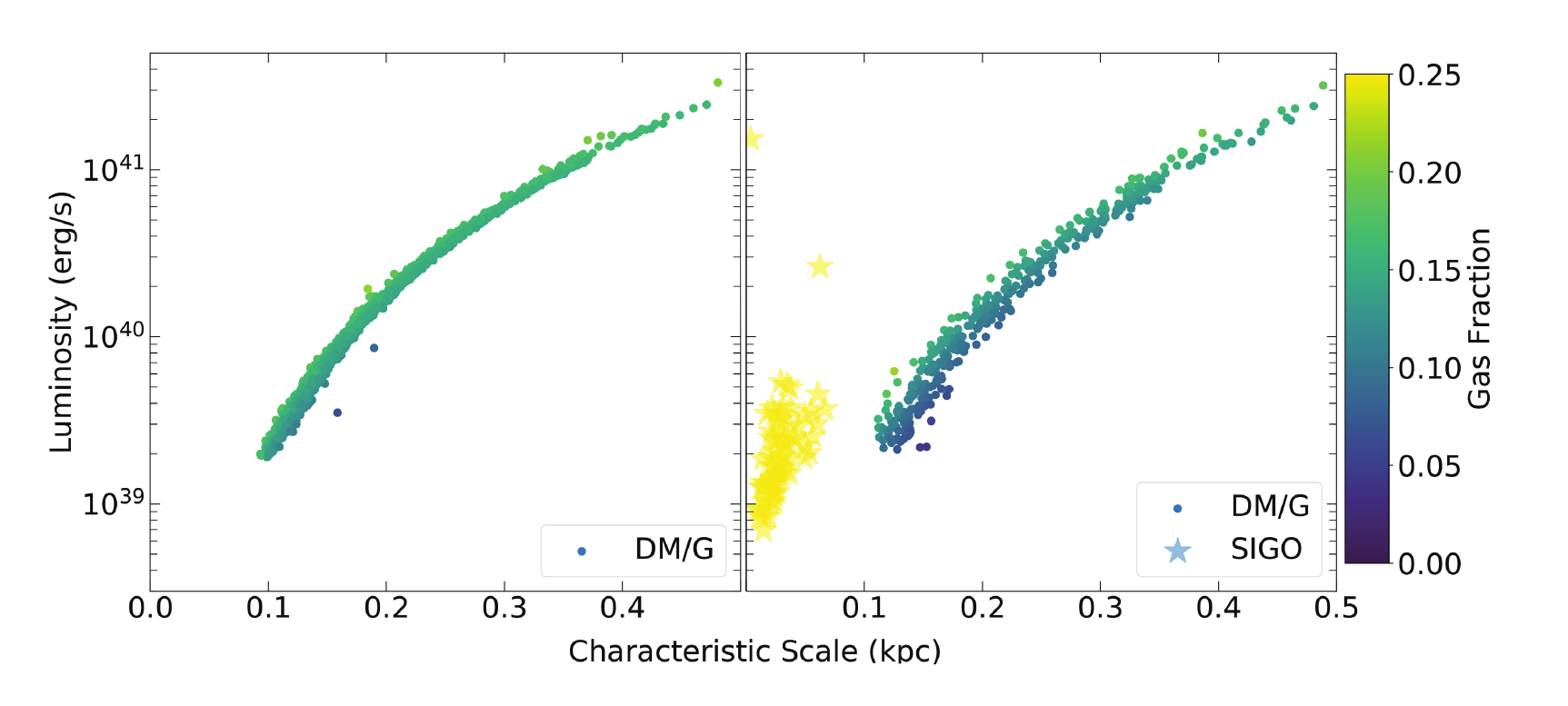}
\end{center}
\caption{Luminosity as a function of the characteristic scale  for $v_{\rm bc} = 0$ (left), $2\sigma_{v_{\rm bc}}$ (right). For the SIGOs, the characteristic scale chosen is the minimum ellipsoid axis. For DM/G objects, we adopt the virial radius. Only star-forming objects with $\rho > \rho_{\rm crit}$ are shown.} 
\label{fig:lumvsradius}
\end{figure*}

\section{SIGO and DM/G Luminosity}
\label{sec:Results}
For the SIGOs and DM/G, with pristine gas, we follow \citet{schaerer03} and consider a star burst model   with no metallicity at redshift $z=20$ (his model ``A''). This  includes Lyman-$\alpha$ lines and the H ionizing photon flux, $Q(H)$. The luminosity is given by
\begin{equation}
L_l [{\rm erg~s}^{-1}] = c_l (1-f_{\rm esc}) Q_i(t) [{\rm s}^{-1}],
\end{equation}
where $Q_i$ is the ionizing photon flux\footnote{Note that following the model of \citet{schaerer03}, $Q_i(t)$ has a linear dependency on the mass of the object.}, $c_l$ is the line emission coefficient for Case B, and $f_{\rm esc}$ is the photon escape fraction. We assume a photon escape fraction of $f_{\rm esc} = 0.5$ and that $10\%$ of the gas mass of a SIGO or a classical object (DM/G) will be converted into stars.

With these relations at hand we estimate the luminosity-mass (Figure \ref{fig:lumvsmass}) and luminosity-radius (Figure \ref{fig:lumvsradius}) relation for the different objects. As it can be clearly seen in these Figures, the SIGOs and DM/G occupy different parts of the parameter space. Moreover, as shown in Figure \ref{fig:lumvsmass}, SIGOs cover the GC mass range.

As expected, and noted previously in the literature \citep[e.g.,][]{Stacy+10,greif11}, the stream velocity suppresses the abundance of DM/G objects and in particular the star-forming ones. In Figure \ref{fig:lumvsradius}, we display the results of the runs with and without stream velocity. The left panel corresponds to the no stream velocity case. Here, there are no SIGOs present and there is a fairly tight relation. With stream velocity, on the right panel, there are less DM/Gs and there is more scatter in the distribution. 

Considering the SIGOs, they occupy a dimmer and more compact part of the parameter space. Since in general star formation would occur in high density peaks that are much smaller than the size of the SIGO, we consider the characteristic scale to be the smallest ellipsoid axis\footnote{Recall that we use $R_{\rm max}$ to calculate the the density threshold, because $R_{\rm max}$ describes the turbulence scale.}(whereas the characteristic scale for the DM/G is simply the virial radius). Indeed, it has been argued that GCs might form as the nuclei of a dwarf galaxy that dissolved \citep[e.g.,][]{searle+78}. Since luminosity is calculated based on the total gas mass of the object and SIGOs tend to be not be very massive, there is a separation in luminosity-mass space. As for luminosity-radius space, the prolate nature of SIGOs gives them a distribution of sizes and they tend to be less luminous in general than the DM/G. 

 The mass and characteristic scale of the SIGOs seems to be consistent with   Little Blue  Dots \citep{elmegreen+17} and the star forming dwarf detected recently by \citet{vanzella+19}. The aforementioned observed objects have been suggested to be GCs progenitors, their similarity to SIGOs is uncanny and may suggest a strong link between high redshift, star-forming SIGOs and GCs progenitors.  Future HST and JWST observations may yield stronger evidence.   

\section{Discussion}
\label{sec:Conclusions}

The supersonically induced gas objects (SIGOs) are expected to exists in patches of the Universe with non-negligible stream velocity  \citep{naoznarayan14,popa,chiou18}. We showed that these gas-rich objects, with little to no DM components, have high enough densities that can give rise to star formation. Thus, the early Universe is predicted to have two classes of star forming objects, the classical ones, i.e., high gas densities within DM halos (DM/G), as well as SIGOs. 

\begin{figure}
\includegraphics[width=0.5\textwidth]{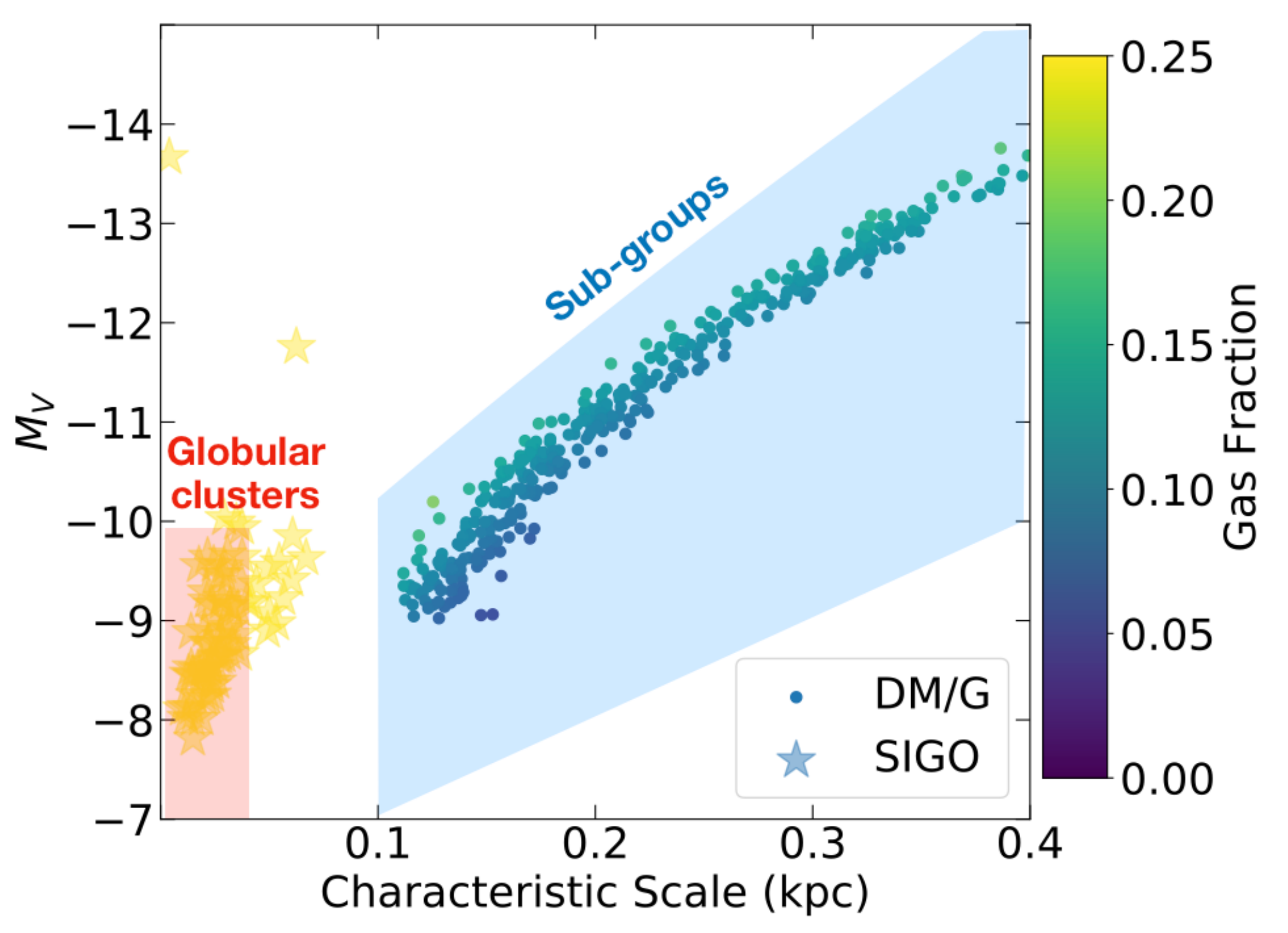}
\caption{The speculated present-day, local, absolute visual magnitude as a function of characteristic scale  of SIGOs and DM/G (see text for details). Over-plotted are object classes in the Local Group from \citet{mcconnachie12}.}
\label{fig:SIGOmag}
\end{figure}

We estimated the luminosity expected from star formation in these objects (both SIGOs and the classical objects). Due to the formation nature of SIGOs, they occupy different parts of the parameter space than the classical DM halos with gas. The SIGOs are dimmer than the classical objects at the same redshift. Note that, while the simulation snapshot here is associated at $z=20$, we expect these objects to continue to form\footnote{Note that $\sim10^6$~M$_\odot$ objects are expected to be fairly common (represent about a $1-\sigma$ fluctuation) at about $z\sim 6$ \citep[e.g.,][]{Naoz+07,fialkov2012,Barkana16}. } and exist (at least before reionization), based on the agreement between the analytical calculations \citep{naoznarayan14} and our simulations \citep{popa,chiou18,chiou+19prep}.. Thus, future JWST observations may be able to disentangle star forming SIGOs from classical objects. 

Moreover, we note that the recently observed Little Blue Dots \citep{elmegreen+17}, which are suggested to be star-forming progenitors of globular clusters, are consistent with with the mass and radius of SIGOs in the simulation. The star-forming dwarf found by \citet{vanzella+19} has also a similar mass and size to our largest SIGOs. There may also be a connection between SIGOs and Giant HII Regions and HII Galaxies \citep{terlevich+18}.  Furthermore, we note that SIGOs that formed little to no stars may be connected to the starless dark HI objects predicted by  \citet{burkhart+16}. Interestingly, the recent discoveries of two galaxies with little to no dark matter \citep{van_Dokkum+18,van_Dokkum+19,Danieli+19}, share a striking resemblance to SIGOs. While the size estimation of these low redshift galaxies is somewhat larger (few kpc) than the SIGOs ($1-100$~pc), we speculate that these objects may be a result, in the local Universe, of a collections or mergers of SIGOs. Moreover, the 10 GCs identified around one of these galaxies \citep{Danieli+19}, are consistent with multiple high densities peaks we have found within our high redshift simulated SIGOs.    

The separation of SIGOs and DM/G in the luminosity-radius parameter space (e.g., Figure \ref{fig:lumvsmass}) highly resembles the magnitude-radius separation parameter space of present-day, local, globular clusters and sub-groups separation \citep[e.g.,][see their figure 6]{mcconnachie12}. Thus, we may speculate on how SIGOs and DM/G objects will be observed today. Assuming a burst-like star formation before reionization ($z=10$), we adopt an initial mass function (IMF) for the objects. In particular, we adopt a top-heavy IMF for the SIGOs\footnote{Following \citet{decressin+07},  we use a piecewise IMF, with low mass end ($0.1\msun < M < 0.8\msun$) is given by a lognormal form and above $0.8\msun$ it is given by a top heavy power law with slope $x=0.55$. Note that using a Salpeter slope for the SIGOs IMF did not significantly affect the results.} following \citet{decressin+07}, and a Salpeter IMF for the DM/G. We then calculate the fraction of spectral types of stars that evolve along the main sequence.  The majority of the stars that survive to present-day then will be G and K type stars, as well as red giants. Given this population, we subtract their various bolometric corrections. We can then roughly estimate each object's visual bolometric magnitude. We also estimate that the observed stellar cluster that formed within the SIGOs corresponding to the highest density peak which is, typically smaller than $R_{\rm max}$. Thus, we adopt the $R_{\rm min}$ for the observed value. Our order of magnitude estimations are presented in Figure \ref{fig:SIGOmag}. We also over-plot the region of the parameter space that is associated with globular clusters (red box) and Andromeda and the Milky Way sub groups (blue area) \citep{mcconnachie12}. Heuristically, the SIGOs are consistent with the absolute visual magnitudes of present-day, local, globular clusters. Although the SIGOS in this simulation only contain primordial gas, we speculate that some self-enrichment or second population formation mechanism (such as pericenter passage of orbits about the nearest DM halo may contribute to the nonzero metallicity in metal-poor GC). Further simulations including explicitly star formation (and the associated metal enrichment) are needed to address this. Nevertheless, the agreement between our rough estimates and the observations is very encouraging.

Finally, these results suggest that if this is the dominant formation mechanism of globular clusters, varying patches (on the order of few tens of Mpcs) in the Universe associated with different coherent $v_{\rm bc}$ values, will have significantly distinct abundances of globular clusters\footnote{ From linear theory, the abundance of SIGOs on face value should follow the abundance of gas poor DM halos \citep[e.g.,][]{naoznarayan14}. Numerically, SIGOs undergo two body relaxation processes and numerically evaporate, \citep[e.g.,][]{popa,chiou18}, thus, detailed abundance studies are challenging. Note that our choice of a larger $\sigma_8$ provides higher power that enables us to overcome some of the numerical challenges and can be viewed as a high-fluctuation patch in the Universe. }. Indeed, about $39\%$ of the Universe contains patches of stream velocity with $\vbc \geq 1\sigvbc$ \citep{tseliakhovich11}. Thus, detailed HST and future JWST observations may allow to disentangle between different formation channels of globular clusters.

\acknowledgments
The authors would like to thank Alice Shapley for leading UCLA GalRead, and leading enlightening discussions, in particular about globular cluster and dwarf galaxy parameter space. We also thank Alice Shapley and Brad Hansen for useful discussions about the observations, and Steve Furlanetto  and Jordon Mirocha for useful discussions about the semi-analytical calculations. We thank the anonymous referee for useful comments that improved the paper. YSC would also like to thank Rick Mebane for useful comments. SN thanks Howard and Astrid Preston for their generous support. BB acknowledges the generous support of the Simons Foundation and discussions with Greg Bryan and Jerry Ostriker. FM is supported by the Program ``Rita Levi Montalcini'' of the Italian MIUR. MV acknowledges support through an MIT RSC award, a Kavli Research Investment Fund, NASA ATP grant NNX17AG29G, and NSF grants AST-1814053 and AST-1814259.
Finally, we thank Volker Springel for granting us access to the {\sc arepo} code.  The simulations presented here were run using Simons Foundation Flatiron Institute computational resources.

\end{document}